\documentclass[pra,showpacs,floatfix,amsmath,amsfonts,twocolumn]{revtex4}

\usepackage{graphics}
\usepackage{epsfig}
\usepackage{color}
\usepackage{eurosym}
\usepackage[normalem]{ulem}
\usepackage[makeroom]{cancel}

\newcommand {\tu}{\tilde{u}}
\newcommand {\tbeta}{\tilde{\beta}}

\newcommand{\PT}{{\cal PT}}

\newcommand{\bM}{\mathbf{M}}

\newcommand{\rev}[1]{\textcolor{black}{#1}}

\begin{document}

\title{Coherent perfect absorber and laser for nonlinear waves in optical waveguide arrays}

\author{Dmitry A. Zezyulin$^1$, Herwig Ott$^2$, and Vladimir V. Konotop$^3$}

\affiliation{$^1$ITMO University, St. Petersburg 197101, Russia\\
$^2$Department of physics and OPTIMAS research center, Technische Universit\"at Kaiserslautern,  Erwin Schr\"odinger Stra{\ss}e,   67663 Kaiserslautern, Germany\\
$^3$Departamento de F\'{i}sica and Centro de F\'{i}sica Te\'orica e Computacional, Faculdade de Ci\^encias, Universidade 	de Lisboa, Campo Grande, Ed. C8   Lisboa 1749-016, Portugal}

\begin{abstract}
A localized non-Hermitian potential can operate as a coherent perfect absorber or as a laser for nonlinear waves. The effect is illustrated for an array of optical waveguides, with the central waveguide being either active or absorbing. The  arrays situated to the left and to the right from the center can have  different characteristics. 
The result is generalized to setups  with the central waveguide carrying additional nonlinear dissipation or gain and to the two-dimensional arrays with embedded one-dimensional absorbing or lasing sub-arrays.
\end{abstract}

\maketitle
 
Spectral singularities and time reversed spectral singularities of non-Hermitian optical potentials, correspond to coherent lasing~\cite{Siegman,longhi2010,Stone2011} and to  coherent perfect absorption (CPA)~\cite{StonePRL,longhi2010,Stone2011,review}. These are special solutions of linear scattering problems  describing either propagation of emitted waves outwards the potential or complete absorption of incident waves. Both lasing and CPA, are not an exclusive property of linear systems, but can also occur when a non-Hermitian potential is nonlinear. This idea was formulated and elaborated in a series of studies~\cite{Mostafa2013,Gupta2013,Mostafa2014,Gupta2014,Argyr}, where one-dimensional non-Hermitian potentials  were considered to bear an additional Kerr nonlinearity in the absorptive or active part. Nonlinear lasing or CPA potentials can be also constructed for two- and three-dimensional scattering problems~\cite{KonZez2018}. In such cases one still deals with absorption or emission of \emph{linear} waves, i.e. waves propagating in linear media, interacting with a nonlinear potential embedded in it. 

Recently, it was shown that the concept of CPA can be extended to \emph{nonlinear} waves~\cite{ZK,matter_absorber}, i.e., waves which propagate in a nonlinear medium and interact with a linear non-Hermitian potential. Moreover, the  CPA of  nonlinear waves was observed experimentally~\cite{matter_absorber} in an atomic  Bose-Einstein condensate (BEC) loaded in a  periodic lattice with one dissipative site from which atoms were eliminated. Unlike CPA of linear waves, the nonlinear CPA  solution (when it is stable) is an attractor and has an additional adjustable parameter, which is the wave amplitude. Therefore, CPA of nonlinear waves  is achievable even more easily than its linear counterpart. On the other hand, the nonlinear CPA does not allow to use such powerful analytical tools, like study of the zeros of the transfer matrix~\cite{Mostafazadeh2009}, i.e., of spectral singularities  and time reversed spectral singularities. Additionally,  the nonlinear CPA might be vulnerable to dynamical  instabilities typical for  nonlinear systems. 
     
In this Letter, we describe CPA and lasing of nonlinear waves in an array of  optical  waveguides with Kerr nonlinearity, where the central waveguide is either active or absorbing. The optical setting allows for further extension of the theory to cases of distinct media at different sides of the central waveguide,  focusing nonlinearities,  nonlinear gain and losses, as well as the absorption and lasing of nonlinear waves in two-dimensional waveguide arrays. 
 
\begin{figure}
	\includegraphics[width=\columnwidth]{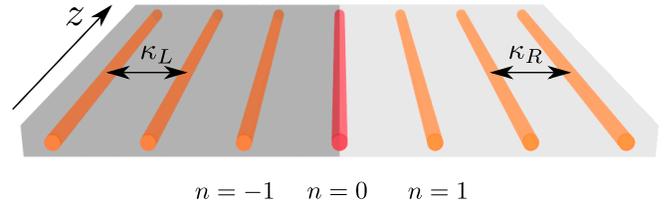}
	\caption{Schematic illustration of the waveguide array which consists of a left (``$L$'') and right (``$R$'') subarrays separated by an absorbing or emitting central waveguide at $n=0$.}
	\label{fig:one}
\end{figure}

We consider two waveguide arrays separated by an active or absorbing waveguide as illustrated in Fig.~\ref{fig:one}.  The waveguides have either positive ($\chi>0$) or negative ($\chi<0$) Kerr nonlinearity. The coupling coefficients in the left  ($\kappa_L$) and right   ($\kappa_R$) arrays
have either equal or different values and   
can be controlled by the dielectric permittivity of the medium between the waveguides.  Without loss of generality, we consider  $\kappa_{L,R}>0$. Light propagation in such an array  is governed  by the discrete nonlinear Schr\"odinger equations
\begin{equation}
\label{DNLS}
\begin{array}{ll}
i\dot{q}_n+\kappa_L\left(q_{n-1} +q_{n+1}\right) +\chi |q_n|^2q_n=0,   \quad  \mbox{for } n\leq -1,
\\
\displaystyle{i\dot{q}_0+\kappa_Lq_{-1} +\kappa_Rq_{1} -i\frac{\gamma}{2} q_0 +\left(\chi-i\frac{\Gamma}{2}\right) |q_0|^2q_0=0, }
\\
i\dot{q}_n+\kappa_R\left(q_{n-1} +q_{n+1}\right) +\chi |q_n|^2q_n=0,\quad   \mbox{for } n\geq 1,
\end{array}
\end{equation} 
where $\dot{q} = dq/dz$, $z$ being the propagation distance, $\gamma$ and $\Gamma$ describe, respectively,  linear and nonlinear gain ($\gamma>0$, $\Gamma>0$) or dissipation ($\gamma<0$, $\Gamma<0$) in the waveguide with $n=0$.   

In the linear case spectral singularities and time-reversed spectral singularities can be determined through the elements of the transfer matrix $\bf M$~\cite{Mostafazadeh2009} {(see also~\cite{matter_absorber,Longhi} where the linear limit of (\ref{DNLS}) was discussed)}. To define it, we consider a stationary wave $q_n=\tu_ne^{i\tbeta z}$ of (\ref{DNLS}) at $\chi= \Gamma=0$, where $\tbeta$ is the propagation constant of the linear solution, and look for the field in the form of the superposition of incident and reflected waves:
$u_n=a_Le^{ik_1n}+b_Le^{-ik_1n}$ for $n\leq 0$ and $u_n=a_Re^{ik_2n}+b_Re^{-ik_2n}$ for $n\geq 0$, where the wavenumbers $k_{1,2}$ are related through the dispersion relations in the left and right subarrays:
\begin{eqnarray}
\label{lin:disp_rel}
\tbeta=2\kappa_L\cos(k_1)=2\kappa_R\cos(k_2).
\end{eqnarray} 
Without loss of generality, we consider $k_{1,2}\in (0,\pi)$. Furthermore, if $k_{1,2}\in(0,\pi/2)$ [$k_{1,2}\in (\pi/2,\pi)$] then $\tbeta>0$ [$\tbeta<0$], and we call the corresponding solutions  slow [fast]  beam. The transfer matrix is computed from the relation $(a_R,b_R)^{\rm T}=\bM (a_L,b_L)^{\rm T}$, where superscript ${\rm T}$ stays for the   transpose matrix. Requiring the continuity of the field at $n=0$, it is  straightforward algebra to compute zeros of the diagonal element $M_{jj}$ ($j=2$ for spectral singularity and $j=1$ for time reversed spectral singularities) of the matrix $\bM$ in terms of the wavevectors $k_{1,2}$ for a given $\gamma$:
\begin{eqnarray}
\label{ss_linear}
 \tbeta(\tan k_1 + \tan k_2) =(-1)^{j}{\gamma}.
\end{eqnarray}
Note that  $\gamma$ is negative for $j=1$ and positive for $j=2$ for all $k_{1,2}$ satisfying the dispersion relation  (\ref{lin:disp_rel}). This   reflects  that the presence of the time-reversed spectral singularity corresponds to the CPA. The respective solution is achieved by setting $b^L=a^R=0$. Similarly, the spectral singularity corresponds to the gain at the central waveguide, and the respective lasing  solution has $a^L=b^R= 0$.

 \rev{Now we consider nonzero values of $\chi$ and $\Gamma$. The idea of the destructive interference of coherent reflected and transmitted waves , which is fundamental for the linear coherent lasing and CPA~\cite{StonePRL,Stone2011,review}, is not applicable for nonlinear media.  Therefore the concept of CPA or lasing for nonlinear waves must be specified. We do this by requiring the nonlinear CPA or lasing to be transformed into their linear counterparts if the field intensity goes to zero. This leads requirements for nonlinear CPA and laser solutions as follows. They  must  (i) be stationary processes; (ii) be observed for monochromatic waves; (iii) be waves propagating respectively inwards and outwards of the central absorbing (or active) waveguide, and
   (iv) have {\em constant amplitude} (we denote it by $\rho$).  Notice that the last requirement is specific to the chosen model and stems  from the linear limit of (\ref{DNLS}). It might require a  modification when CPA or lasing of nonlinear waves is considered in more sophisticated models.} Nonlinear solutions satisfying the above requirements can be searched as
   \begin{equation}
 \label{nonlin_solut}
 q_n=e^{i\beta z}
 \left\{
 \begin{array}{ll}
 \rho e^{\pm ik_1n} & \mbox{for } n\leq 0, 
 \\
 \rho e^{\mp ik_2n} & \mbox{for }  n\geq 0,
 \end{array}
 \right.
 \end{equation}
where  he propagation constant of nonlinear solution reads 
 \begin{eqnarray}
 \label{beta_nl}
 \beta=2\kappa_L\cos k_1+\chi\rho^2=2\kappa_R\cos k_2 +\chi\rho^2,
 \end{eqnarray} 
 and  upper (lower) signs stand for CPA (lasing) solutions. Formula (\ref{nonlin_solut}) is a solution of (\ref{DNLS}) for $k_{1,2}$ related to $\gamma$ by 
 \begin{eqnarray}
 \label{ss_nonlinear}
\tbeta(\tan k_1 + \tan k_2) = \mp({\gamma} + \Gamma \rho^2). 
 \end{eqnarray}
At  $\Gamma=0$,  Eq.~(\ref{ss_nonlinear}) coincides with  expression (\ref{ss_linear}) valid for the linear case.  
The presence of the coefficient $\Gamma$ enriches the behavior of the system, since in this case the amplitude $\rho$ of the absorbed (or emitted) nonlinear wave  enters expression (\ref{ss_nonlinear}) explicitly mediating values of the coefficients $\gamma$ and $\Gamma$ for which CPA or lasing take place for given $k_1$ and $k_2$. In particular, the competition between the coefficients $\gamma$ and $\Gamma$ in (\ref{ss_nonlinear}) enables the situation  when lasing (absorption) of nonlinear waves can be achieved in the presence of linear absorption (gain) and nonlinear gain (absorption), provided the amplitude of the nonlinear wave is large enough, i.e. at $\rho^2>|\gamma/\Gamma|$.
 
 Since in a discrete medium the group velocity of excitations is bounded, there exists a largest value of the loss (gain) for which CPA (lasing) is possible. In the case when properties of the left and right media are different, there  also appears the lower limit for loss (gain) supporting the CPA (lasing). In both, linear and nonlinear media, a necessary condition for existence of constant amplitude solutions takes the form
\begin{eqnarray}
\label{bounds}
\sqrt{2|\kappa_L^2-\kappa_R^2|}<|\gamma+\Gamma \rho^2|\leq {2}( \kappa_L+\kappa_R).
\end{eqnarray} 
Thus if the linear gain (loss) is absent, $\gamma=0$, the amplitude of coherently absorbed (emitted) waves is bounded  from above and from below.   
  
Nonlinear systems can be vulnerable to dynamical instabilities. This rises a question about the stability of the obtained nonlinear CPA and laser waves. Solution (\ref{nonlin_solut})  has the form of counterpropagating    uniform backgrounds whose stability is a necessary requirement for the stability of the nonlinear CPA and laser solutions. This  implies that in the case of $\chi>0$ ($\chi<0$) only fast (slow) CPA or lasing of nonlinear waves can be observed (see e.g.~\cite{matter_absorber, Kevrek}). In either of these cases, like in the linear limit, the left hand side of (\ref{ss_nonlinear}) is positive for all $k_{1,2}$. 

\begin{figure}[!t]
 	\begin{center}
 		\includegraphics[width=0.85\columnwidth]{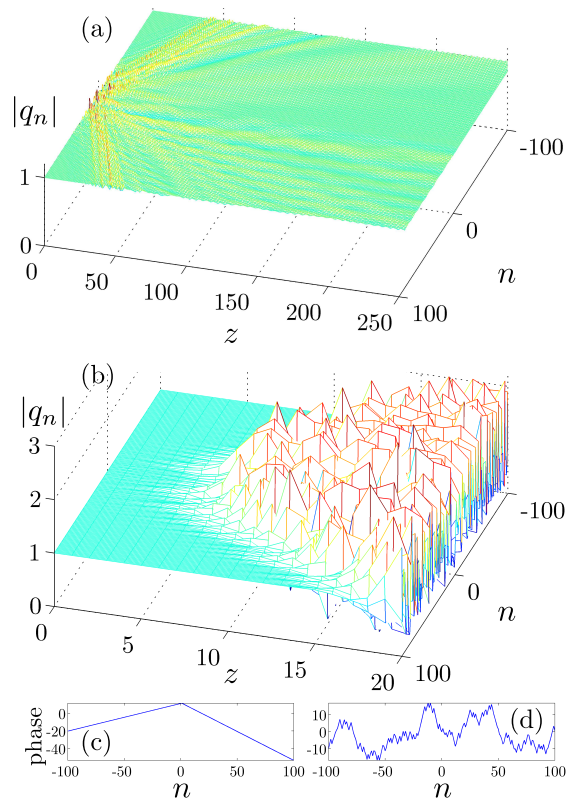}%
 		\caption{(a) Restoration of the nonlinear CPA solution (slow beam) from a distribution with a strong initial perturbation. Here $\kappa_L=1$, $\kappa_R=1.2$, $k_1\approx 0.31$, $k_2\approx 0.66$, $\gamma\approx -2.08$, $\Gamma=0$, $\chi=-1$, and $\rho=1$; (b) Instability of the coexisting fast beam with the same parameters except for wavevectors being $k_1 \approx 2.83$ and $k_2 \approx 2.49$. (c,d) Unwrapped radian arguments (phases) of the fields in the end of the simulated  propagations. \label{fig:dyn01}}
 	\end{center}
 \end{figure}

Stability of the uniform backgrounds does not guarantee yet the stability of CPA or laser solutions, since the latter may be unstable with respect to perturbations spatially localized in the vicinity of the central waveguide. The robustness of the  CPA and laser solutions for each particular set of parameters can be checked numerically by simulating the propagation of the solution with initial perturbation. We have performed a series of such simulations and observed that the CPA and laser solutions can indeed be stable for various combinations of the system parameters. Moreover, stable nonlinear CPA and laser solutions feature attractor-like behavior emerging from significantly distorted input beams, which is typical for dissipative systems. An example is presented in Fig.~\ref{fig:dyn01}(a) where an initially perturbed slow beam (with sufficiently strong random perturbation localized around the central waveguide)  recovers the nearly uniform intensity distribution. The CPA nature of the resulting solution can be observed in the $\Lambda-$shaped distribution of the phases of the field shown in Fig.~\ref{fig:dyn01}(c). The latter is composed of  two nearly straight lines whose slopes are determined by wavevectors $k_1$ and $k_2$, being a signature of CPA  in accordance with Eq.~(\ref{nonlin_solut}). In Fig.~\ref{fig:dyn01}(a) we present the simplest CPA solution with the absorption implemented using only the linear losses $\gamma$ with zero nonlinear gain or losses $\Gamma\ne 0$. At the same time, similar stable CPA solutions can be also constructed for nonzero (positive and negative) values of $\Gamma$. The latter provides some  additional flexibility allowing to manipulate  parameters of the CPA solution (such as $\rho$, $k_{1,2}$), or to tune other parameters of the absorber (i.e., $\gamma$). In particular, for sufficiently strong nonlinear absorption, which corresponds to large negative  $\Gamma$, the CPA solution  remains valid even if $\gamma\geq 0$, i.e., even if there is the  linear gain acting in the system. 
 
For the fast beam, even a weak perturbation seeded in the vicinity of the central waveguide rapidly triggers strong modulational instability as illustrated in Fig.~\ref{fig:dyn01}(b). The phase distribution of the resulting field loses the characteristic $\Lambda$-shaped    pattern    already on a short propagation distance [Fig.~\ref{fig:dyn01}(d)], meaning that no coherent absorption takes place.   
   
\begin{figure}[!t]
	\begin{center}
				\includegraphics[width=0.85\columnwidth]{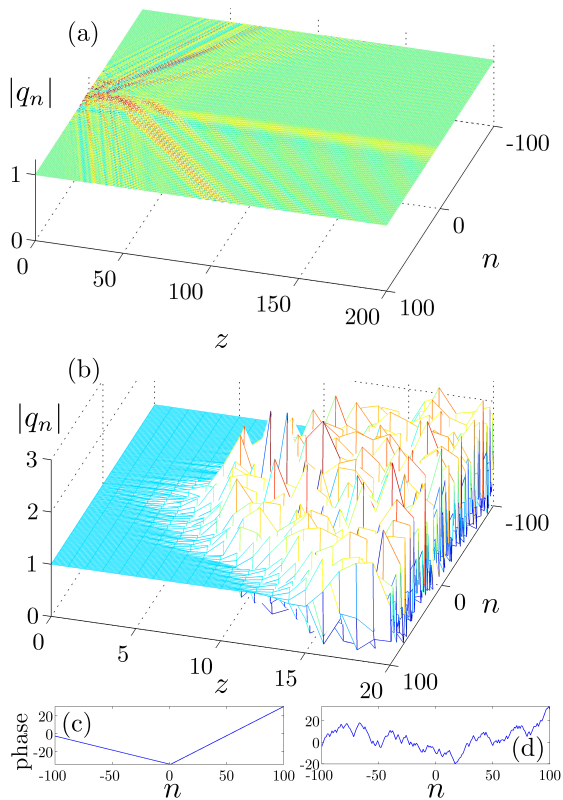}
		\caption{(a) Propagation  of a stable nonlinear laser   solution (slow beam) from a distribution with an  initial perturbation. Here $\kappa_L=1$, $\kappa_R=1.2$, $k_1\approx 0.31$, $k_2\approx 0.66$, $\gamma\approx 3.08$, $\Gamma=-1$, $\rho=1$, and  $\chi=-1$; (b) Instability of the coexisting  fast beam with the same parameters except for wavenumbers being $k_1\approx 2.83$ and $k_2 \approx 2.49$. (c,d) Unwrapped radian arguments of the fields  in the end of the simulated propagations. \label{fig:dyn02}}
	\end{center}
\end{figure}

Similar results for a laser solution supported by linear gain and nonlinear losses at the central site are presented in Fig.~\ref{fig:dyn02}. The stable slow beam solution maintains its nearly constant-amplitude intensity in spite of an initial perturbation [Fig.~\ref{fig:dyn02}(a)] and features the distinctive V-shaped  phase pattern [Fig.~\ref{fig:dyn02}(c)] which indicates clearly    the lasing nature of the solution. Unstable fast beam solution rapidly develops the modulational instability [Fig.~\ref{fig:dyn02}(b)] and loses the distinctive phase distribution [Fig.~\ref{fig:dyn02}(d)]. In general, while the exhaustive analysis of stability for CPA and laser solutions is left beyond the scope of this Letter, our numerics allow to infer that solutions with smaller $|\gamma|$ are in general more robust, and the CPA  solutions are more robust than their laser counterparts.

Solution~(\ref{nonlin_solut}) can be  generalized to the case where the left and right arrays are have different Kerr nonlinearities (i.e., $\chi = \chi_L$ for $n\leq -1$ and   $\chi =\chi_R$ for $n\leq 1$). Now (\ref{beta_nl}) is replaced by 
\begin{equation}
\beta=2\kappa_L\cos k_1 + \chi_L\rho^2  = 2\kappa_R\cos k_2 + \chi_R\rho^2.
\end{equation}
 If the nonlinearity of the central waveguide is chosen as $\chi_0 = (\chi_L+\chi_R)/2$,
 solution (\ref{nonlin_solut}) remains valid provided
$
\kappa_L\sin k_1 + \kappa_R\sin k_2 = \mp ({\gamma} + \Gamma \rho^2)/2,
$
where  upper and lower signs  again correspond to the CPA and laser solutions, respectively.

\begin{figure}
	\includegraphics[width=\columnwidth]{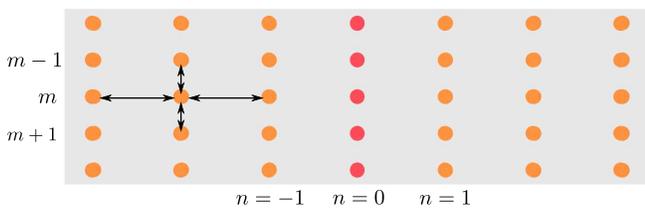}
	\caption{Schematic illustration of the 2D waveguide array with the uniform nearest-neighbor coupling  and the  embedded absorbing or lasing layer at $n=0$.}
	\label{fig:2D}
\end{figure}

The above consideration can be generalized to the two-dimensional (2D) waveguide arrays  with an embedded  absorbing or lasing ``layer'' as schematically shown in Fig.~\ref{fig:2D}. This situation can be consided   as a discrete analogue and a nonlinear generalization of previously considered surface  CPAs for linear waves  \cite{2D1,2D2,2D3,KonZez2018}. Now, the field in each waveguide $q_{n,m}$ is described by two subscripts $n$ and $m$. For fixed $m$ and running $n$, one recovers the one-dimensional case considered above, while changing $m$  establishes the second dimension. Respectively, the absorbing or lasing layer is situated at $n=0$ with $m$ running through all integers. 
We restrict the consideration to the case of equal media at both sides of the absorbing or lasing layer. The  2D  array is described by the system 
\begin{equation}
\label{DNLS2D}
\begin{array}{l}
i\dot{q}_{n,m}+\kappa \Delta_2 q_{n,m} +\chi |q_{n,m}|^2q_{n,m}=0,   \quad  \mbox{for } n \ne 0,  \\[1mm]
i\dot{q}_{0,m}+\kappa \Delta_2 q_{0,m} \displaystyle{-i\frac{\gamma}{2} q_{0,m} +\left(\chi-i\frac{\Gamma}{2}\right) |q_{0,m}|^2q_{0,m}=0}, 
\end{array}
\end{equation} 
where  
$\Delta_2q_{n,m} = q_{n-1,m} +q_{n+1,m} +q_{n,m-1} +q_{n,m+1}$.

\begin{figure}
	\includegraphics[width=\columnwidth]{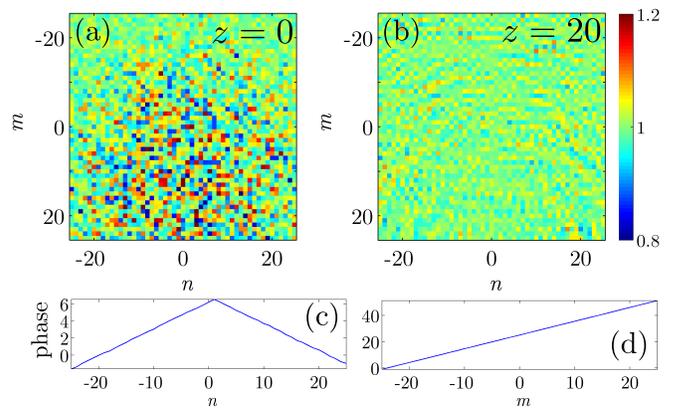}	
	\caption{Stable propagation of the initially perturbed 2D CPA solution. (a) and (b) show the intensities $|q_{n,m}|$  at $z=0$ and $z=20$, respectively. Both (a) and (b) correspond to the same colorbar shown on the right; (c) and (d) show the unwrapped radian arguments along the first dimension (for $m=0$) and along the second dimension (for $n=0$), for $z=20$.     Here $\rho=1$,  $\kappa=1$,  $k\approx 0.31$,   $\nu=\pi/3$, $\gamma\approx -1.24$, $\Gamma=0$, $\chi=-1$.}
	\label{fig:2Ddyn}
\end{figure}

The 2D  nonlinear CPA or lasing solution can be found as
 \begin{equation}
\label{nonlin_solut2D}
q_{n,m}(z)=\rho e^{i\beta z+i\nu m\mp ik |n|  }
\end{equation}
where new wavevector $\nu\in (-\pi, \pi)$ is a free parameter   which tunes the angle of incidence (angle of emission) of the  nonlinear wave with respect to the absorbing or lasing layer.  The nonlinear propagation constant reads $\beta = 2\kappa (\cos k + \cos \nu) + \chi \rho^2$, and the following condition must be satisfied: $4\kappa \sin k = \mp (\gamma + \Gamma\rho^2)$. As above, upper and lower signs correspond to CPA and laser, respectively.   An example of stable propagation of the 2D   nonlinear CPA  solution is shown in Fig.~\ref{fig:2Ddyn}.  The initial distribution is chosen in the form (\ref{nonlin_solut2D}) with an additional random perturbation whose maximal amplitude was about $10\%$ of the average solution amplitude. During the propagation, the perturbation decays, and the  solution recovers the nearly uniform amplitude.   Lower panels in Fig.~\ref{fig:2Ddyn} highlight different phase distributions of the propagating field along the first ($n$) and the second ($m$) dimensions: in the former case the phase pattern is conditioned by the energy flows towards the absorbing layer, whereas in the latter case the phase increases monotonously according to the prescribed value of wavenumber $\nu=\pi/3$ in Fig.~\ref{fig:dyn02}.

To conclude, we have demonstrated that optical waveguide arrays represent a platform for the realization of a CPA or a lasing of nonlinear waves. 
The constant-amplitude solutions studied herein  represent the generalization of the concept of (time-reversed) spectral singularities on the case of a nonlinear guiding medium. \rev{Nonlinear CPA and lasing solutions are attractors and can be used for rectification of the phase gradient and for equalization of the amplitudes of input beams. Indeed, small initial deviations from the exact nonlineaar solution will be damped along the propagation. Yet another related feature is that by reducing the input field amplitudes it is possible to achieve the regime of linear CPA and lasing where the conditions of the destructive interference will be satisfied automatically. In the meantime a number of interesting related questions are left open. Among them we mention an eventual possibility of nonlinear analogs of a $\PT$-symmetric CPA-laser~\cite{longhi2010}, of splitting of self-dual spectral singularities~\cite{split}, as well as of a coherent virtual  absorbers~\cite{Baranov}. The implementation of the latter one with linear arrays of waveguides~\cite{Longhi}, as well as resemblance of their properties with nonlinear light concentrators introduced in~\cite{BKA},   indicate  a possibility of developing nonlinear virtual absorbers by combining a linear defect with Kerr nonlinearity.} 


\section*{Funding information} 
Ministry of Education and Science of Russian Federation (Megagrant No. 14.Y26.31.0015, Grant 08-08).    
The German Research Foundation DFG within the SFB/TR 185 OSCAR.

\end{document}